\pdfoutput=1
\documentclass{aa}
\usepackage{graphicx}
\usepackage{txfonts}
\usepackage{color}
\usepackage{tabularx} 
\usepackage{adjustbox}
\usepackage{ulem}
\usepackage{mydblfloatfix}
\usepackage{xcolor}

\newcommand{\be}{\begin{equation}}
\newcommand{\ee}{\end{equation}}
\newcommand{\bea}{\begin{eqnarray}}
\newcommand{\eea}{\end{eqnarray}}

\newcommand{\sinc}{\mbox{sinc}}

\newcommand{\ha}{H$\alpha$}

\begin{document}

\title{ALMA observations of the variability of the quiet Sun at millimeter wavelengths
}
\author{A. Nindos \inst{1}
\and S. Patsourakos \inst{1}
\and  C.E. Alissandrakis \inst{1}
\and T.S. Bastian \inst{2}}
\institute{Physics Department, University of Ioannina, Ioannina GR-45110,
Greece\\
\email{anindos@uoi.gr}
\and
National Radio Astronomy Observatory, 520 Edgemont Road, Charlottesville VA
22903, USA}

\date{Received: Accepted:}

 
  \abstract
 {}
{We address the variability of the quiet solar chromosphere at 1.26\,mm and 3\,mm
with a focus on the study of spatially resolved oscillations and transient
brightenings, i.e. small, weak  events of energy release. Both phenomena may
have a bearing on the heating of the chromosphere.}
{We used Atacama Large Millimeter/submillimeter Array (ALMA) observations of
the quiet Sun at 1.26\,mm and 3\,mm. The spatial and temporal resolution of the
data were $1-2\arcsec$ and 1\,s, respectively. The concatenation of light curves
from different scans yielded a frequency resolution in spectral power of
0.5-0.6\,mHz. 
 At 1.26\,mm, in addition to power spectra of the original data, 
we degraded the
images to the spatial resolution of the 3\,mm images and used fields of view
of equal area for both data sets. The detection of transient brightenings was
made after the effect of oscillations was removed.}
{At both frequencies we  detected p-mode oscillations in the range 3.6-4.4 mHz. The corrections for spatial resolution and
field of view at 1.26\,mm decreased the rms of the oscillations by factors of 1.6 and 1.1, 
respectively. In the
corrected data sets, the oscillations at 1.26\,mm and 3\,mm showed
brightness temperature
fluctuations of $\sim$1-7-1.8\% with respect to 
the average quiet Sun, corresponding to
137 and 107\,K, respectively. We detected 77
transient brightenings at 1.26\,mm and 115 at 3\,mm. Although their majority 
occurred in cell interior, the occurrence rate per unit area of the 1.26\,mm 
events was higher than that of the 3\,mm events and this conclusion does not
change if we take into account differences in spatial resolution and noise
levels. The energy associated with the transient brightenings ranged from
$1.8 \times 10^{23}$ to $1.1 \times 10^{26}$\,erg and from $7.2 \times 10^{23}$ to
$1.7 \times 10^{26}$\,erg for the 1.26\,mm and 3\,mm events, respectively. The 
corresponding power-law indices of the energy distribution were 1.64 and 1.73.
We also found that ALMA bright network structures corresponded to dark mottles/spicules seen in broadband \ha\ images from the GONG network. }
{The fluctuations associated with the p-mode oscillations represented a 
fraction of 0.55-0.68 of the full power spectrum. Their energy density at 1.26\,mm
was $3 \times 10^{-2}$ erg cm$^{-3}$. The computed low-end energy of the 1.26\,mm
transient brightenings is among the smallest ever reported, irrespective of
the wavelength of observation. Although the occurrence rate per unit area
of the 1.26\,mm transient brightenings was higher than that of the 3\,mm events
their power per unit area is smaller probably due to the detection of many weak
1.26\,mm events.}


   \maketitle
%

\section{Introduction}

The chromosphere is one of the most complex layers of the solar atmosphere
(e.g. see Carlsson et al. 2019 for a recent review).  It is the layer that
marks the transition from a plasma dominated regime to a magnetic field 
dominated regime. Since it is the interface between the photosphere and the
higher, much hotter layers of the solar atmosphere, the chromosphere cannot be 
ignored in any attempt to address the heating of the overlying corona. 
In addition, the
chromosphere
is rich in
dynamic phenomena on different
spatial and  temporal scales (e.g. see the reviews by Schmieder 2007,
Tsiropoula et al. 2012, and Shimizu 2015, and  references therein), not only in active regions but also in so-called quiet  regions. 

Prominent among them in the quiet Sun are oscillations and wave phenomena as well as
small-scale weak transient activity. Both oscillations and transient brightenings may 
have a bearing on the heating of the chromosphere. Oscillations and brightenings, together
with non-periodic time variations, turbulence and instrumental noise contribute to the
observed time variation at any point in the chromosphere. Hence, care must be taken in order to separate all these components.

Oscillations
are ubiquitous in the
quiet chromosphere (e.g. see the review by Jess et al. 2015 and
references therein), with
periods from three to five
minutes; they reflect the penetration of photospheric p-modes into the
chromosphere (e.g. see de Wijn et al. 2009 and references
therein). The highly structured nature of the chromosphere (even of
the quiet one) contributes to the appearance of a variety of wave and
oscillatory phenomena, including reflections, interferences, mode
conversions and even shocks (see Wedemeyer-B\"{o}hm et al. 2009). 

Small-scale episodes of energy release are also ubiquitous in the quiet 
chromosphere (e.g. see Henriques et al. 2016 and references therein). They
have been detected in a wide range
of wavelengths, from soft X-rays to microwaves,
and their observational signatures exhibit significant diversity.
The
low-end detection limit of these events is determined by the capabilities 
of the instrument. In the literature there is no unique name for them; in this 
article we adopt the term ``transient brightenings''. 

The interpretation of chromospheric observations at wavelengths that range
from the ultraviolet to infrared is not easy due to departures from local
thermodynamic equilibrium (LTE). The situation is better at millimeter
wavelengths; such emission originates  in
the chromosphere, and in the quiet 
Sun 
is due to
the free-free mechanism under LTE (e.g. see Shibasaki et al.
2011; Wedemeyer et al. 2016). Therefore, the source function is Planckian and 
the observed brightness temperature is directly linked to the electron 
temperature via the radiative transfer equation.

With the advent of the Atacama Large Millimeter/submillimeter Array (ALMA),
observations of the chromosphere at millimeter wavelengths with unprecedented
spatial resolution (a few seconds of arc or less), temporal resolution (1-2
s) and sensitivity have been accumulating (see Loukitcheva 2019 for a review 
of first results). High spatial resolution observations of the quiet Sun with 
ALMA include Shimojo et al. (2017a, 2020); Bastian et al. (2017); Yokoyama et 
al. (2018); Nindos et al. (2018); Jafarzadeh et al. (2019); Loukitcheva et al. (2019);
Wedemeyer et al. (2020).

The first high resolution (about 10\arcsec) observations of millimeter 
wavelength oscillations was reported by White et al. (2006) and Loukitcheva
et al. (2006) who used data from the Berkeley-Illinois-Maryland Array (BIMA)
at 3.5 mm. Although their observations could not clearly separate the network 
from cell interiors, they reported periods of 5 min and longer in the network
while periods in the intra-network were shorter (about 3 min). The first
detection of quiet Sun millimiter oscillations with ALMA was done by 
Patsourakos et al. (2020; hereafter referred to as Paper I) who reported 
spatially resolved oscillations, in both network and cell interior, at 3\,mm 
with frequencies of $4.2\pm1.7$ mHz and rms variation of 55 to 75 K.
Although the time intervals of the observations were only 10-minute long,
p-mode peaks were detected in the power spectra of the original time series. 
In order to accurately measure the oscillation amplitude, they de-trended the 
data by fitting a third-degree polynomial, which suppressed slowly varying fluctuations. 

Jafarzadeh et al. (2021) searched for oscillatory phenomena in 10 ALMA datasets
obtained at 1.26\,mm or 3\,mm which corresponded to observations of regions with 
different concentrations of magnetic flux. 
They found
oscillatory power at frequencies from 3 to 5 mHz 
in 
only two
quiet 
regions, actually the ones used in this article, while lower frequencies dominated the power spectra from regions 
associated with stronger magnetic fields. Guevara-G\'{o}mez et al. (2021)
reported high-frequency oscillations (periods in the range of 66-110 s) at 3 
mm associated with small bright features in a plage/enhanced network region.

Detection of individual weak transient events using ALMA observations at 3\,mm
has been reported by Shimojo et al. (2017b; plasmoid ejection from an X-ray
bright point) and Yokoyama et al. (2018; jet-like activity). The first systematic study
of transient brightenings in ALMA 3\,mm quiet Sun observations was done by 
Nindos et al. (2020; hereafter referred to as Paper II) who used images of
six fields of view each one observed for 10 min (same data as in Paper I). After they removed oscillations
they detected 184 events with brightness temperatures from 70 K to more than 
500 K above background. The typical duration of the events was about 50\,s,
their thermal energies were between $1.5 \times 10^{24}$ erg and $10^{26}$ 
erg, and their frequency distribution as function of energy was a power law 
with an index of 1.67. 

Eklund et al. (2020) analyzed another dataset of ALMA 3\,mm observations of the
quiet Sun and detected 552 events in the course of their 40-min observations.
Their events exhibited excess brightness temperatures from more than 400 K to 
about 1200 K (typical values were between 450 K and 750 K). Their typical 
duration was from about 55 to 125 s. Eklund et al. (2020) reported that the
properties of several of their events were consistent with the signatures
of propagating shock waves (see also the simulations presented in Eklund et al.
2021). 
Finally, we mention that transient brightenings in 
3\,mm ALMA observations of a small active region have been reported by da Silva 
Santos et al. (2020). Their events corresponded to much higher excess 
brightness temperatures (up to about 14200 K) than the quiet Sun events of 
Paper II.

In this article we present a systematic study of the millimeter wavelength
variability in a quiet Sun region that was observed by ALMA at 1.26\,mm and 3\,mm, 
in which we clearly separated oscillations and transients.
After we present the data and an overall view of the region we studied (Section
2) we focus on the analysis of oscillations (Section 3) and transient 
brightenings (Section 4). We present conclusions in Section 5.

\begin{figure*}
\centering
\includegraphics[width=\hsize]{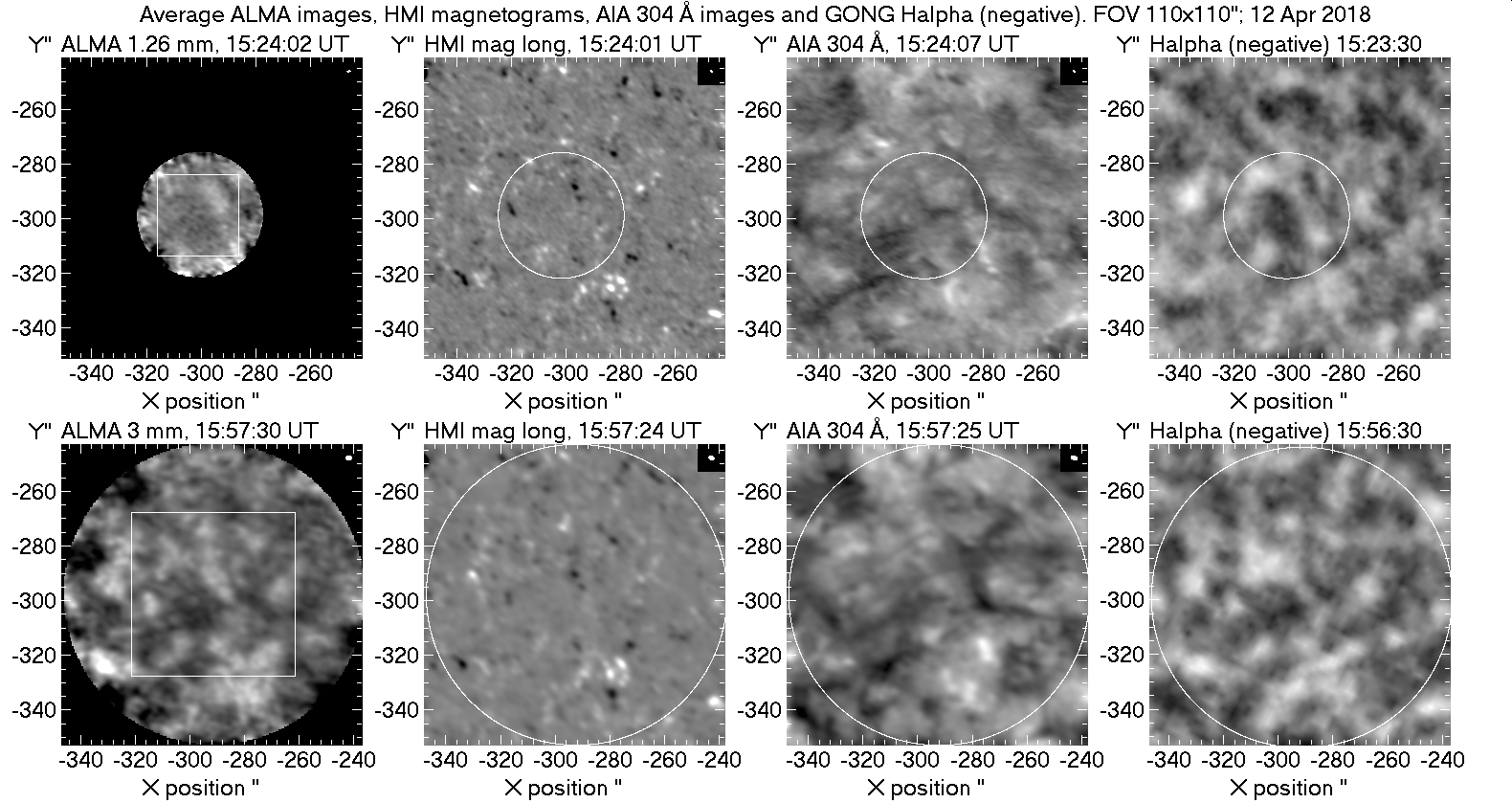}
\caption{Average images during ALMA scan 8 in Band 6  (top, left) and 
during scan 1 in Band 3 (bottom, left), together with the corresponding 
HMI magnetograms (saturated at $\pm50$\,G), as well as AIA 304 \AA\ and \ha\ 
negative images  from GONG. The insert at the top right corner of the images shows 
the ALMA resolution. The squares in the ALMA images show the regions analyzed 
for oscillations, the circles in the other images mark the ALMA field of view. 
All images are oriented with the celestial north up.}
\label{fig:images}
\end{figure*}

\section{Data reduction and overall view of the observations}

Our observations took place on 12 April 2018  and employed two spectral bands of ALMA: Band 6 at 1.26 mm 
(239\,GHz)
and Band 3 at 3\,mm (100 GHz). The same region was observed sequentially, first in Band 6 and then in Band 3; it was centered at heliocentric coordinates 
of about -132\arcsec, -398\arcsec\ ($\mu \approx 0.9$).

Band 6 observations consisted of 8 scans  (Table \ref{ALMAobs}), from 13:59:00.39 to 15:25:02.69 UT. For scans 1-4 and 5-8 there was an interruption of about 2.5 min for phase calibration between consecutive scans; scans 4 and 5 were separated by about 20 min due to amplitude calibration. Each scan had a duration of about 7.9 min, with the exception of scans 4 and 8 which had a duration of about 2 min.

Band 3 observations consisted of 6 scans, from 15:52:30.49 to 17:16:00.69. For scans 1-3 and 4-6 there was again an interruption  of about 2.6 min between consecutive scans, while scans 3 and 4 were separated by about 20 min. Each scan had a duration of about 9.9 min, with the exception of scans 3 and 6 which had a duration of about 6.9 min.

From the Joint ALMA Observatory we received calibrated visibilities 
(their calibration employed the scheme described by Shimojo et al. 2017a) as 
well as average CLEAN images. No full-disk component was added to these images. 
Furthermore, their comparison with simultaneous 1600 \AA\ data taken with the
Atmospheric Imaging Assembly (AIA; Lemen et al. 2012) aboard Solar Dynamics 
Observatory (SDO) revealed that the ALMA pointing was not accurate. 
The origin of the problem was that a correction for general relativistic 
deflection was not turned off, as should have been the case for all solar 
observations. The magnitude of the effect depends on the nominal pointing 
offset relative to disk center (N. Phillips, private communication). We 
computed the actual pointing by cross-correlating the ALMA images with the 
1600 \AA\ images and verified that the position offset we established was
consistent with the general-relativistic radial offset.

Subsequently, for each scan we re-computed average CLEAN images 
synthesized from the entire observing interval as well 1\,s snapshot images
(see Nindos et al. 2018 for details). Both the average and snapshot images 
were computed after four rounds of phase self-calibration and one round of
amplitude self-calibration. The self-calibration improved the quality of
the image cubes and reduced
the jitter appearing in the movies of the
images made prior to the application of self-calibration. The improvement, 
judged by the amplitude and root mean square (rms) values of the jitter as well 
as the width of the autocorrelation function of the images, was more 
significant in the Band 6 data, apparently due to the stronger influence of seeing at higher frequencies. Finally, all snapshot images were combined with
contemporaneous full-disk ALMA images in order to recover large spatial scales
that are not available to the interferometer, and obtain absolute brightness 
temperature calibration. We note that the correction of the brightness temperature at the solar disk center deduced by  Alissandrakis et al. (2020) for ALMA Band 6 applies only to full disk images and not to the interferometric images, 
whose flux calibration is established via the ALMA Calibrator Device
(see Shimojo et al. 2017a for details).

The image cadence was 1\,s, which provided a temporal sampling two
times better than that of our March 2017 observations
(Nindos et al. 2018).  In addition,  for the present data
  set we achieved spatial resolution of 1.6\arcsec\ by  0.7\arcsec\ for Band
6 and 2.7\arcsec\ by 1.7\arcsec\ for Band 3; this
was
superior to that of our 2017 observations (3-4\arcsec\ for Band 3).
The pixel size and diameter of the field of view of the images were 0.25\arcsec\ 
and 46.5$\arcsec$ for Band 6 and 0.5\arcsec\ and 110$\arcsec$ for Band 3. 

\begin{table}[h]
\begin{center}
\caption{ALMA quiet Sun observations on 12 April 2018}
\label{ALMAobs}
\begin{tabular}{cccc}
\hline 
Scan& Wavength& UT range& Duration\\
   &    mm         &               &   min  \\
\hline
 1  & 1.26 & 13:59:00.39 - 14:06:57.11 & 7.95 \\
 2  & 1.26 & 14:09:29.62 - 14:17:26.34 & 7.95 \\
 3  & 1.26 & 14:19:59.53 - 14:27:56.25 & 7.95 \\
 4  & 1.26 & 14:30:29.43 - 14:32:28.61 & 1.99 \\
 5  & 1.26 & 14:51:22.71 - 14:59:19.43 & 7.95 \\
 6  & 1.26 & 15:01:55.73 - 15:09:52.45 & 7.95 \\
 7  & 1.26 & 15:12:29.09 - 15:20:25.81 & 7.95 \\
 8  & 1.26 & 15:23:03.51 - 15:25:02.69 & 1.99 \\
\hline 
 1  & 3.00 & 15:52:30.49 - 16:02:27.40 &  9.95\\
 2  & 3.00 & 16:05:05.24 -  16:15:02.15 & 9.95 \\
 3  & 3.00 & 16:17:39.89 - 16:24:36.01 &  6.93 \\
 4  & 3.00 & 16:43:54.20 - 16:53:51.11 &  9.95 \\
 5  & 3.00 & 16:56:29.33 - 17:06:26.24 &  9.95 \\
 6  & 3.00 & 17:09:04.57 - 17:16:00.69 &  6.93 \\
\hline 
\end{tabular}
\end{center}
\end{table}



Fig.~\ref{fig:images} shows average images during the last scan in ALMA Band 
6 (2\,min duration) and the first scan in Band 3 (9.9\,min duration). In the 
same figure we give the corresponding Helioseismic and Magnetic Imager 
(HMI) aboard SDO magnetograms (saturated at $\pm50$\,G), as well as AIA 304 
\AA\ images, both smoothed with the ALMA beam; we also give single frames of 
the best nearest \ha\ images from the Global Oscillation Network Group 
(GONG), displayed as negatives. 

We note that the observed region was very quiet,
so that the network (contrary to our 2017 observations) is not very well discernible 
in the magnetograms. Still, 
most peaks in the mm-wavelength images, at 3\,mm in particular, are 
well associated with network elements, of both positive and negative polarity. 
A large ($\sim25$\arcsec\ diameter) supergranular cell, centered at about 
$x=-298\arcsec, y=-305\arcsec$ (Fig.~\ref{fig:images}), covers a large part 
of the 1.26\,cm field of view. 
Finally, the negative \ha\ images (centered at line center) from the 
GONG network show 
a strong similarity to the ALMA images, at 3\,mm in particular; 
this is probably because GONG images are broad-band, hence they are dominated 
by the absorption of dark mottles/spicules, which are located above network 
elements. Note that Molnar et al. (2019) have reported a good correlation
between \ha\ core width maps and ALMA brightness temperature at 3 mm.
Average values for network and cell interior brightness temperatures 
for the ALMA data sets analyzed in this article have been given by 
Alissandrakis et al. (2020).

\section{Analysis of Oscillations}

\label{sec:osc} 
\begin{figure*}
\centering
\includegraphics[width=.8\hsize]{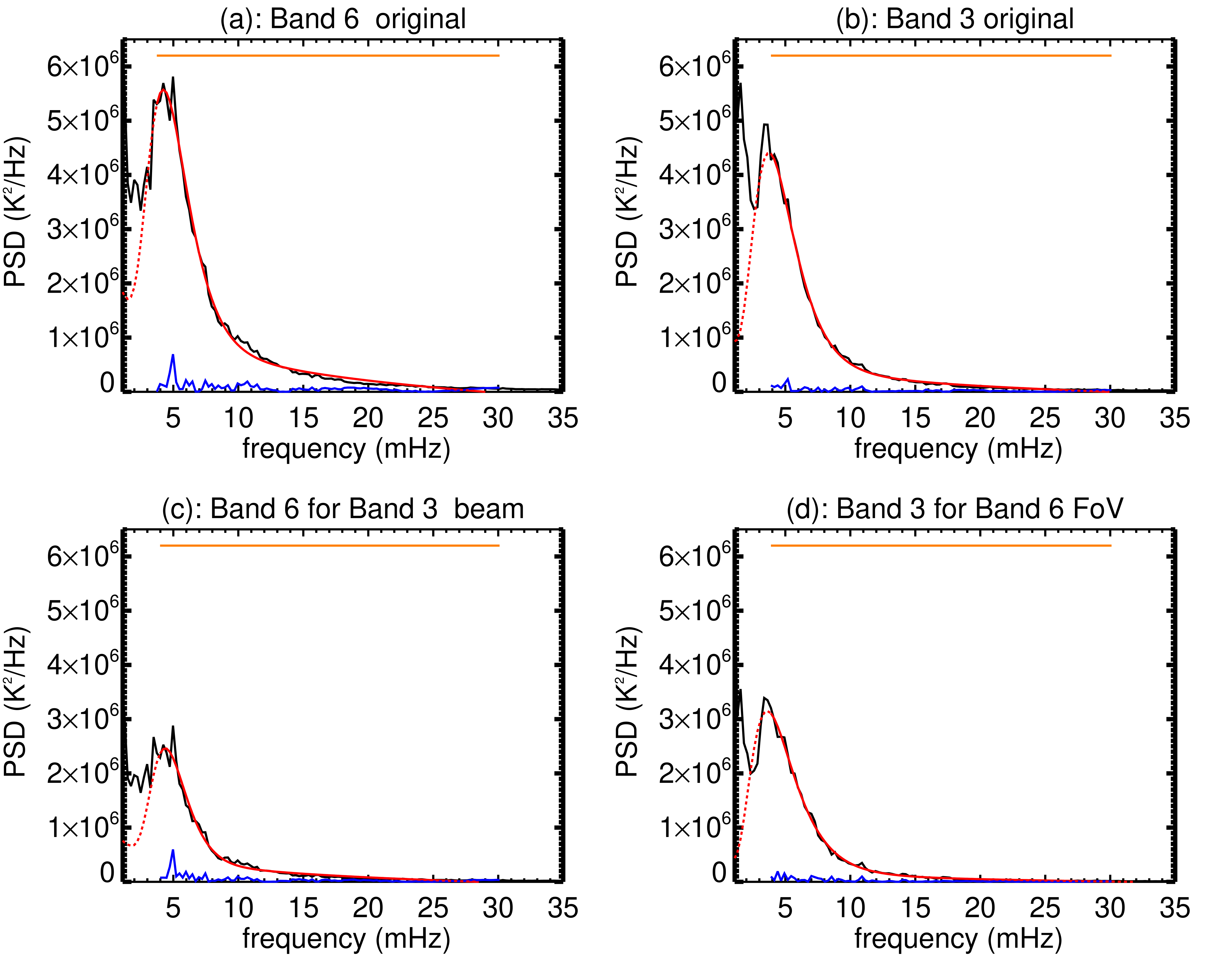}
\caption{Spatially averaged PSDs for
the series discussed in Section \ref{sec:osc}:
Band 6 original  data (panel a), Band 3 original  data  (panel b),
Band 6 
data degraded to the resolution of Band 3
(panel c), and Band 6 for Band 3 FoV (panel d).
The spectra of panels (c) and (d) are  appropriate 
for meaningful comparisons between
Band 6 and Band 3. Black solid lines represent the observed PSDs; the 
horizontal orange line corresponds to the frequency range employed in the  fits;
red solid curves show the fit
of the observed PSDs and red dashed lines show how the best fit 
extends outside the frequency range used in the fit;
the solid blue lines
show the absolute value of the fit
residuals.}
\label{fig:psd}
\end{figure*}

A first report on the oscillations in our April 2018 data has been presented
by Jafarzadeh et al. (2021). We note, however, that these authors used pointing
information for the ALMA images which was different from the one we use in this
article.
We   selected 
 120 pixel $\times$ 120 pixel fields of view (FoV)
around the center of each image to avoid artifacts resulting
from the primary beam correction toward the edges of
each FoV. Given the different pixel sizes for Band 6 and Band 3, the corresponding FoVs
were 30 \arcsec  $\times$ 30\arcsec  and 60 \arcsec  $\times$ 60\arcsec\
respectively. The selected regions are marked by the squares in the ALMA images of Fig. \ref{fig:images}.

Since there were large interruptions between scans 4 and 5 of Band 6 observations, we 
did not attempt to analyze the entire time series. Instead, we split the data 
in
two extended blocks containing 
scans 1-4 and 5-8, 
including both actual observations and gaps between scans,
with a duration per block of $\approx$ 33 min. 
%
Likewise for Band 3  observations, given the large interruption between scans 3 and 4, 
we  made
two extended blocks, containing  
scans 1-3 and 4-6 with a total 
duration per block of $\approx$ 27 min. 
We refer
to these series as Band 6 original and Band 3 original, respectively. The frequency resolution of the
Band 6 and Band 3 observations for the blocks resulting from the concatenation
of consecutive scans discussed above is 0.5 and 0.6 mHz, respectively. 

\begin{table*}[t]
\begin{center}
\small
\caption{Parameters of p-mode oscillations from ALMA observations at 3 and 1.26\,mm
from this study (rows 2-5) and 
from Paper I (row 1). }
\begin{tabular}{lccccccc}
\hline
                     & p-mode     &p-mode             &Full spectrum&Full spectrum   & p-mode/full &Peak frequency&Peak FWHM   \\
Data                 &   RMS (K)  &RMS/$T{_b,qs}$     &RMS (K)      &RMS/$T{_b,qs}$     &   ratio   & (mHz) & (mHz)   \\
\hline
3.00\,mm; Paper I &$~~66\pm~~6$  &  $0.009~\, \pm 0.0008$ & $105 \pm 14$  & $0.0140 \pm  0.0019$ & $0.64 \pm 0.04 $  &$4.2 \pm 0.07$ &  $ 3.7 \pm 0.07 $   \\
3.00\,mm Band 3 original  & $152\pm1$ & $0.0208\pm0.00002$ & $254\pm5~\,$ & $0.0347\pm0.0006$& $0.59\pm0.01$ &$3.8 \pm 0.01$  & $ 3.7 \pm 0.08 $   \\
1.26\,mm Band 6  original  & $175\pm~1$ & $0.0275\pm0.0001$ & $283\pm7~\,$ & $0.0446\pm0.001$& $0.62\pm0.02$ &$4.3 \pm 0.01$ &  $ 3.6 \pm  0.04 $ \\
3.00\,mm Band 3  for Band 6 FoV & $137\pm4$ & $0.0187\pm0.0005$ & $200\pm4~\,$ & $0.0273\pm0.0005$& $0.68\pm0.04$ & $3.6 \pm 0.21$ & $ 3.9 \pm  0.05 $  \\
1.26\,mm Band 6  for Band 3 beam    & $109\pm~1$ & $0.0171\pm0.0001$ & $195\pm10~\,$ & $0.0307\pm0.0015$& $0.55\pm0.04$ &
$4.4 \pm $ 0.02 &  $ 3.2 \pm 0.16$ \\
\hline
\label{tab:psd}
\end{tabular}
\end{center}
\end{table*}

In order to properly  compare the Band 3 and Band 6 temporal variations, 
differences in both the beam size and the 
FoV for these two bands  should be taken into account. 
Therefore, the superior spatial resolution Band 6 images were first deconvolved with the Band 6 beam, and 
then convolved with the Band 3 beam. This essentially yields Band 6 observations of the same spatial resolution as
those of Band 3, and we thereby refer to the series from this operation as Band 6  for Band 3 beam.
Next, in order to take into account the larger Band 3 FoV, we trimmed the 
Band 3 original  FoV into a FoV which is identical to the Band 6 FoV (i.e. 30 \arcsec  $\times$ 30\arcsec), and we hereby refer
to results from this series as Band 3  for Band 6 FoV. Therefore,
the Band 6 and Band 3 series produced from these calculations, correspond 
to observations with the same spatial resolution and FoV, therefore they can be
subjected to meaningful comparisons. 

The light curves at each pixel of the four series described above (hereafter referred 
to as Band 6 original, Band 3 original,
Band 6  for Band 3 beam, and Band 3 for Band 6 FoV),
and per given block (i.e., scans 1-4 and 5-8 for Band 6 and scans 1-3 and 3-6 for Band 3) 
were then submitted
to Fourier transform in order  to calculate the corresponding Power Spectral
Density (PSD), defined as:
\be
\mathrm{PSD}(\nu)=\frac{1}{T}P(\nu),
\ee
where $\nu$ is the frequency, 
$T$ is the duration of the time series and $P(\nu)$ is the 
power spectrum
of the observed light curve 
(Equation \ref{psp} in 
the Appendix). The average value of 
the
light curve of each individual pixel was subtracted prior
to the PSD calculation  and zeroes were added to fill the blanks in the concatenated light curves.
Next, for each block we calculated the spatially-averaged 
PSD over the corresponding FoV, and finally averaged the PSD
for the two blocks considered for each band. The above procedure allowed to take 
into account the longest possible light curves and, 
at the same time, to derive the most representative spectra for the entire sequence.  

The resulting PSDs 
for the four series 
are displayed with the black solid lines in Fig. \ref{fig:psd}. In all cases the p-mode peak  
in the vicinity of $\approx 3.6-4.4$ mHz,  is standing well above the spectral background.
The longer duration of the present observations compared to the observations
reported in Paper I ($\approx 10$ min) allowed to better track the rollover of spectral power
toward lower frequencies; it was thus not necessary to de-trend the data, as we had done in Paper I. The presence of sidelobes around the peak in all  spectra  is also obvious, and this
is more prominent in Band 6. The introduction 
of gaps in our light curves introduces these sidelobes in the resulting power spectra and also
influences their amplitude. In the calculations of peak amplitudes
we are reporting throughout this paper, we corrected for the decrease
in power due to the gaps.  A detailed discussion of the impact of the gaps on the resulting power spectra and correction factors is
given in the Appendix.  

The effect of degrading the spatial resolution of the Band 6 observations to
that of the Band 3 observations is quite pronounced and leads to a significant 
decrease 
(factor $\approx$ 2 at spectral peak) of the power (compare panels (a) and (c) of Fig. \ref{fig:psd}).
This is because the decrease of spatial resolution reduces the short wavelength components of the oscillations.
Moreover, trimming the Band 3 FoV into the Band 6  FOV,  leads to further decrease of the  power by a factor of 1.4 at spectral peak
(compare panels (b) and (d) of Figure
\ref{fig:psd}). This time the reduction happens because the limitation of the FoV reduces the amplitude of the long wavelength components of the p-modes.

To proceed into a more quantitative analysis  and in order to compute the total power 
of the chromospheric p-modes we followed
Paper I and fitted the observed  PSDs with an analytical function consisting 
of the sum of a log-normal function (that is, Gaussian of $\ln (\nu)$, $\nu$ being the frequency) 
and a linear function of $\ln (\nu)$, meant to describe the p-mode (value and frequency
of peak and spectral width) and  
the spectral background, respectively: 
\be
\mathrm{PSD}(\nu)=a_{0}+a_{1} \ln \nu+a_{2}\,\exp \left(-\frac{{{(\ln \nu-\ln a_{3})}}^{2}}{2{a_{4}}^2}\right),
\label{fitpsd}
\ee
where $a_{0}-a_{4}$ correspond to the free parameters of the fitting function.
The function was applied to frequency ranges of the
PSDs marked
with horizontal orange lines in
Fig. \ref{fig:psd} and  these were deduced as follows:
We first considered  ``search'' intervals of potential lower and upper 
frequencies, 
in the 2-4\,mHz and 30-50\,mHz range
respectively.
Next, for each combination of lower and upper frequency that could be drawn from these intervals, we performed
the PSD fitting and calculated the corresponding
${\chi}^{2}$. The lower and upper frequency employed in our analysis
was the one leading to the smallest ${\chi}^{2}$.  Experimentation showed that
considering different ``search'' intervals
for mainly the lower frequency led 
to undesirable results, since the corresponding  fits were not anymore able
to track the spectral peaks. 

In Fig. \ref{fig:psd} we plot the PSD fits
with red solid lines. 
Generally speaking, the employed fitting function
does a decent job in reproducing the observed PSDs,
as it could be also judged by the relatively small
absolute residuals (blue solid lines) not exceeding $\approx 20 \%$.

From the log-normal part of the fitting function 
we calculated the root-mean-square (rms)  associated
with the p-mode (see Equation 3 in Paper I) while the integral of the PSD over the full 
spectrum, as per Parseval's theorem, supplies the rms of the brightenss temperature, 
$T_{b}$, fluctuations for any  temporal variation, oscillatory 
and non-oscillatory, of the corresponding light curves.

In Table \ref{tab:psd} we present results from the  fits
of our Band 6 and Band 3 observations corresponding to the four considered series.
For comparison, we also present the pertinent average values derived from 
the Band 3 observations of Paper I.
Columns 2-8 give the p-mode rms, the ratio of p-mode rms to that of the average
quiet Sun $T_{b}$, the rms of the full spectrum, the ratio of the full spectrum rms to the quiet Sun $T_{b}$ from Alissandrakis et al. (2020), the ratio of the p-mode rms to that of the full spectrum (that is, the ratio of columns 3 and 5), the frequency of p-mode peak, and the full width at half maximum (FWHM) of the p-mode 
peak.
The quoted errors represent the half value of the absolute difference of any given derived quantity between the two blocks of our observations (that is,
scans 1-4 and 5-8 in Band 6 and scans 1-3 and 4-6 in Band 3).
On the basis of the results presented in Table  \ref{tab:psd} we can 
make the following remarks:
\begin{itemize}
\item the fractional p-mode rms (RMS/$T{_b,qs}$), which essentially gives the relative amplitude of 
the p-mode, is similar ($0.17-0.18$) for Band 6 and Band 3;
\item original Band 6 observations, when degraded to the 
resolution of 
Band 3, 
exhibit a significant decrease in the rms by a factor of 1.6 which is somehow smaller
than the corresponding factor in resolution (i.e., $\approx$  2);
\item varying the employed FoV for Band 3 observations (full versus matching that of Band 6)
gives rise to a difference of the rms of about 1.1;  
\item in both bands the p-mode rms corresponds to a significant fraction of the full spectrum rms in the range 0.55-0.68;
\item Band 3 observations of Paper I exhibit smaller p-mode rms and fractional rms with respect to
the observations of this work by a factor of $\approx$ 2.1, apparently due to the inferior resolution of the data set used in Paper I;
\item the frequencies of the p-mode peak lie in the range [3.6-4.4] mHz; 
the Band 6 peak frequencies exceed those of Band 3 but 
the differences are within the spectral resolution of the analyzed power spectra;
\item Band 6 and Band 3 observations of this work and of Paper I exhibit FWHM in the range [3.2-3.9] mHz. The difference of FWHM in the 
Band 6 and 3 original time series falls within the calculated uncertainties.
However, the FWHM in the ``Band 3 for Band 6 FoV'' time series is larger than 
the FWHM in the ``Band 6 for Band 3 beam'' time series, since their difference
exceeds the relevant uncertainties.
\end{itemize}

\begin{table*}[t]
\begin{center}
\caption{Statistics of the 1.26\,mm and 3\,mm transient brightenings. Values in parentheses give the number of events per (cm$^2$\,s), multiplied by a factor of 10$^{-22}$.}
\begin{tabular}{lccccc}
\hline
Data set & ALMA TBs & 304 \AA\ TBs & 1600 \AA\ TBs & ALMA TBs with         & ALMA TBs with \\
        &          &              &               & 304 \AA\ counterparts & 1600 \AA\ counterparts \\
\hline
April 2018, 1.26\,mm & 77       & 101     & 385    & 0 & 12 \\
                 & (5.16)   & (6.76)  & (25.71) & (0.00)  & (0.80)    \\
April 2018, 3\,mm & 115      & 160     & 472    & 10 & 10 \\
                 & (2.39)   & (3.33)  & (9.84) &  (0.21)  & (0.21)   \\
March 2017, 3\,mm & 184      & 199     & 633    & 18 & 14 \\
(Paper II)       & (4.81)   & (5.19)  & (16.60) & (0.47)   & (0.37)   \\
\hline
\label{tab:stat}
\end{tabular}
\end{center}
\end{table*}

From the p-mode rms we  can
make an estimate
of the p-mode energy density, $\epsilon$:
\begin{equation}
\epsilon=\rho {\delta v}^2, 
\label{eq:epsilon} 
\end{equation} 
where
$\rho$  is the mass density and $\delta v$  is the velocity amplitude.
For linear sound waves it is:
\begin{equation}
\frac{\delta T}{T}=(\gamma-1)\frac{\delta v}{c_s},
\label{eq:dt}
\end{equation}
where ${\delta T}$ (equal to $\sqrt{2}$ times 
the temperature rms) is the temperature amplitude of the p-mode, $T$ is the average temperature, ${c_s}$ is the sound speed 
and  $\gamma=5/3$. 

Substituting in Equations \ref{eq:epsilon} and \ref{eq:dt}
the Band 6 fractional rms for Band 6 original data from Table \ref{tab:psd} (this series corresponds
to highest fractional rms)
and 
$\rho$ from the Fontenla et al. (1993) FAL C model corresponding to the average quiet 
Sun temperature in Band 6 at disk center from Alissandrakis et al. (2020)
we obtained
an energy density 
of  $\approx 3 \times {10}^{-2} \mathrm{erg\,{cm}^{-3}} $.
Using observations of absorption lines formed at heights
from about 130 to about 1000 km, Canfield and Musman (1973)
found that the p-mode energy density decreases with
height from  $2 \times {10}^{2}$ erg cm$^{-3}$
at 130 km to  to $4 \times {10}^{-1}$ erg cm$^{-3}$ at
a height of 1000 km. Alissandrakis et al. (2020) 
used full-disk ALMA images to detect the solar limb and derived  
an emission height of $2400 \pm 1700$ km for Band 6, which is
above the heights scanned by Canfield and  Musman (1973).
This implies that the decreasing trend in the p-mode energy
density with height extends above 1000 km.


\section{Transient brightenings}
\label{sec:trans}

\subsection{Detection of transient brightenings}

For the detection of transient brightenings we followed the procedure
described in Paper II. Briefly, we removed the effect of oscillations 
(see Section \ref{sec:osc}) from the light curve of each pixel.
At both ALMA wavelengths we were able to significantly suppress the 
oscillatory power (by factors of 13 to 16 in Band 3, and by factors of 11 to 
15 in Band 6). The maximum residual oscillatory power was also similar at both 
wavelengths; its highest values were 42 K$^2$/Hz in Band 3 and 37 K$^2$/Hz in 
Band 6. These results are similar to those reported in Paper II. As in Paper 
II, the criteria we used for event identification were: (i) at least four 
consecutive points in each ``corrected'' time profile should have intensity 
exceeding a 2.5$\sigma$ threshold above the mean intensity of the light curve, 
(ii) such behavior should also be shared by a beam-size patch of adjacent 
pixels (18 0.5$\arcsec$-pixels for Band 3 and 18 0.25$\arcsec$-pixels for Band 6), 
and (iii) the time profiles of the selected pixels should peak within $\pm$2 min. 
At both frequencies the variation of the threshold parameters affects the results 
in manners similar to those described in Paper II.

We only used scans 1-2 and 4-5  from the Band 3 data and scans 1-3 and 5-7 
from the Band 6 data, i.e. we did not take into account  scans 3 and 6 in 
Band 3 and scans 4 and 8 in Band 6 because their duration was shorter than 
the periods of oscillations (see Section 2). To compensate for the  smaller
field of view of the images compared to the field of view of the images used
in Paper II, we submitted all of their pixels to our algorithm for the
detection  of transient brightenings. Pixels toward the edges of the
field of view could be susceptible to artifacts
introduced by the primary beam correction. However, the number of events
that we detected at the outer 30-pixel ring of each field of view was
rather limited (seven events in Band 3 and six events  in Band 6) and their
properties were indistinguishable from the properties of the rest of
the events. Furthermore, it is obvious from Fig. 1 that much of the
structures imaged in the ALMA images extend beyond their field of view
and the problem is more serious in Band 6.  To this end, we also
searched for transient brightenings in the extended regions (square
fields of view of $128\arcsec \times 128\arcsec$ and  $64\arcsec
\times 64\arcsec$ in Bands 3 and 6, respectively) covered by the
self-calibrated CLEAN'ed images prior to the application of the primary
beam  correction. The possible detection of transient events beyond
the circular fields of view dictated by the primary beam would
increase our  statistical sample, although the photometric properties
of such events would not be accurate. However, we were not able to
recover any more events in these areas.

We used the method described above to detect transient brightenings in 304 
\AA\ and 1600 \AA\ AIA data (same regions and times as the ALMA images). For 
this task, we first convolved the AIA images with the pertinent ALMA beam. 
Due to the inferior cadence of the AIA images (12\,s and 24\,s for 304\,\AA\ and 1600\,\AA, respectively) we reduced the number
of consecutive time profile points exceeding the 2.5$\sigma$ threshold 
from four to one. 

\subsection{Properties of transient brightenings}
 
The numbers of transient brightenings that we detected at both ALMA frequencies
as well as in AIA data appear in Table \ref{tab:stat} together with the 
numbers of events per unit area and time (values in parentheses). For 
comparison we also give the relevant results from Paper II. The numbers 
reported in Table \ref{tab:stat} are uniformly distributed among the time
intervals of the ALMA scans with the exception of the 1.26\,mm events during 
scan 7 where only three events were identified, probably due to inferior 
seeing conditions. 

\begin{table*}
\begin{center}
\caption{Location, range of maximum intensity, area, and duration of the 1.26\,mm 
and 3\,mm transient brightenings}
\begin{tabular}{lccc}
\hline
Parameter & April 2018, 1.26\,mm & April 2018, 3\,mm & March 2017, 3\,mm \\
          &                  &                  & (Paper II) \\
\hline
Network events      & 48\% & 73\% & 68\% \\
Range of maximum intensity (K) & 44-449 & 65-511  & 71-504 \\
Power law index of max. intensity & $1.93 \pm 0.06$ & $2.15 \pm 0.04$ & $2.10 \pm 0.03$ \\
Mean area (Mm$^2$)  & $5.2 \pm 1.8$ & $9.3 \pm 2.1$ & $12.3 \pm 3.4$ \\
Power law index of area distribution & $2.30 \pm 0.03$ & $2.69 \pm 0.04$ & $2.73 \pm 0.02$ \\
Mean duration (s) & $50.7 \pm 6.1$ & $49.7 \pm 5.2$ & $51.1 \pm 6.5$ \\
Power law index of duration distribution & $2.36 \pm 0.04$ & $2.31 \pm 0.03$ & $2.35 \pm 0.02$ \\
\hline
\label{tab:prop}
\end{tabular}
\end{center}
\end{table*}

\begin{figure*}[b]
\centering
\includegraphics[width=0.6\textwidth]{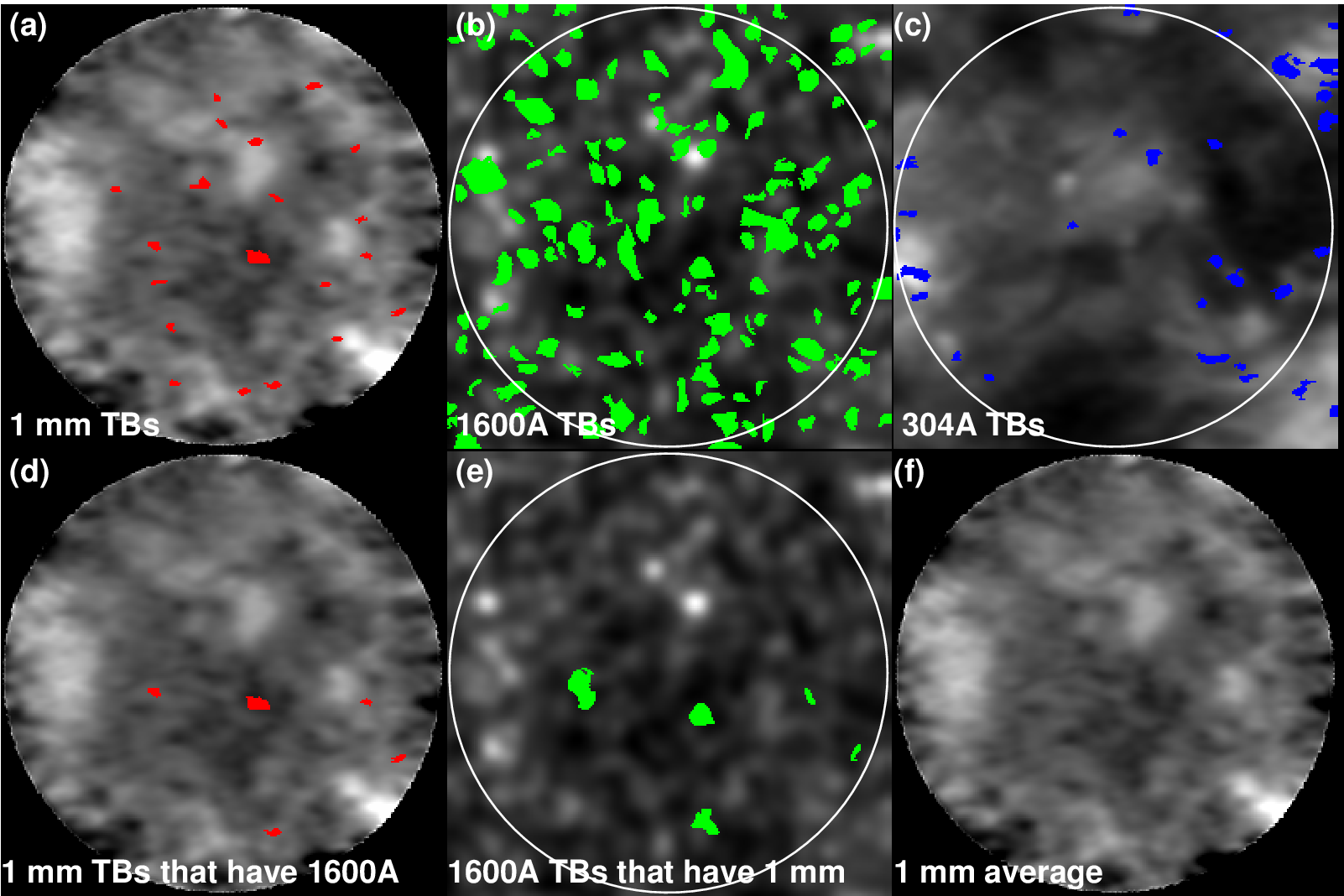}
\caption{Display (overlayed on 
the pertinent average images) of the pixels that were associated with transient bightenings in ALMA 1.26\,mm (red), AIA 1600 \AA\ (green), and AIA 304 \AA\ (blue)  for scan 1 data. In the images of the top row, we mark all events at each wavelength. In panels (d) and (e) of the bottom row we mark only the paired 
events, that is, those that were detected both at 1.26\,mm and 1600 \AA, respectively. Panel (f) shows the average 1.26\,mm image. The white circles denote the ALMA field of view.}
\label{fig:col1}
\end{figure*}

It is interesting that the occurrence rate per unit area of the 1.26\,mm events is 
higher than that of the 3\,mm events both from the data set presented in this 
article and from the data set analyzed in Paper II (factors of 2.16 and 1.07, 
respectively). For the proper interpretation of this result one needs to take
into account the differences both in noise level and spatial resolution between
the Band 3 and Band 6 data. The evaluation of noise was done using the
prescription described by Shimojo et al. (2017a); see also Yokoyama et al. 
(2018). Briefly, in both Bands 3 and 6, ALMA receivers are equipped with systems that 
are sensitive to linear orthogonal polarizations; X and Y. The subtraction of 
simultaneous snapshot images from the XX and YY data is expected to be zero in 
the quiet Sun and any deviation from zero can be attributed to noise. We 
measured noise levels of 10.5\,K and 21.2\,K in Bands 3 and 6, respectively.

We also degraded the resolution of images from selected Band 6 scans to 
the 3\,mm resolution (see also Section \ref{sec:osc}). The resulting images were 
searched for transient brightenings. We did find
smaller number of events
but again their occurrence rate per unit area remained higher (a factor of 
1.40) than that of the Band 3 events. It appears that the very quiet area 
imaged in Band 
6 (the interior of a supergranular cell covers much of the 1.26\,mm field of view; 
see section 2) may have helped in the identification of weak events that  would
otherwise have been obscured by the relatively stronger emissions in the Band 
3 data. This interpretation is consistent with the higher occurrence rate per 
unit area of the AIA data registered with the Band 6 data than that of the AIA 
data registered with the Band 3 data (see Table \ref{tab:stat}).  

The number of 304 \AA\ events is higher than the number of ALMA events at a 
given wavelength and even more so is the number of 1600 \AA\ events (see Table
\ref{tab:stat}). As in Paper II, paired events (that is, ALMA events with AIA
counterparts, either at 304 \AA\ or 1600 \AA) were identified by requesting 
that the ALMA and AIA events have one or more common pixels and their time 
profiles peak within 15 s. 

The number of paired ALMA-AIA events appears in columns five and six of Table 
\ref{tab:stat}. As in Paper II, we did not find any event with conspicuous 
signatures in three wavelengths (that is, ALMA, either 1.26\,mm or 3\,mm, 304 \AA, 
and 1600 \AA). The percentage of 3\,mm events with AIA counterparts is about 9\%,
which is similar to the percentage found in Paper II. On the other hand, we
found no 1.26\,mm events with 304 \AA\ counterparts although 15\% of the 1.26 mm events were
paired with 1600 \AA\ events. The complete absence of 304 \AA\ counterparts to
the 1.26\,mm events is not a result of chance. A simple calculation reveals that
the number of chance concurrences between 1.26\,mm and 1600 \AA\ events should be
between less than two to more than three events. 

\begin{figure*}
\centering
\includegraphics[width=1.0\textwidth]{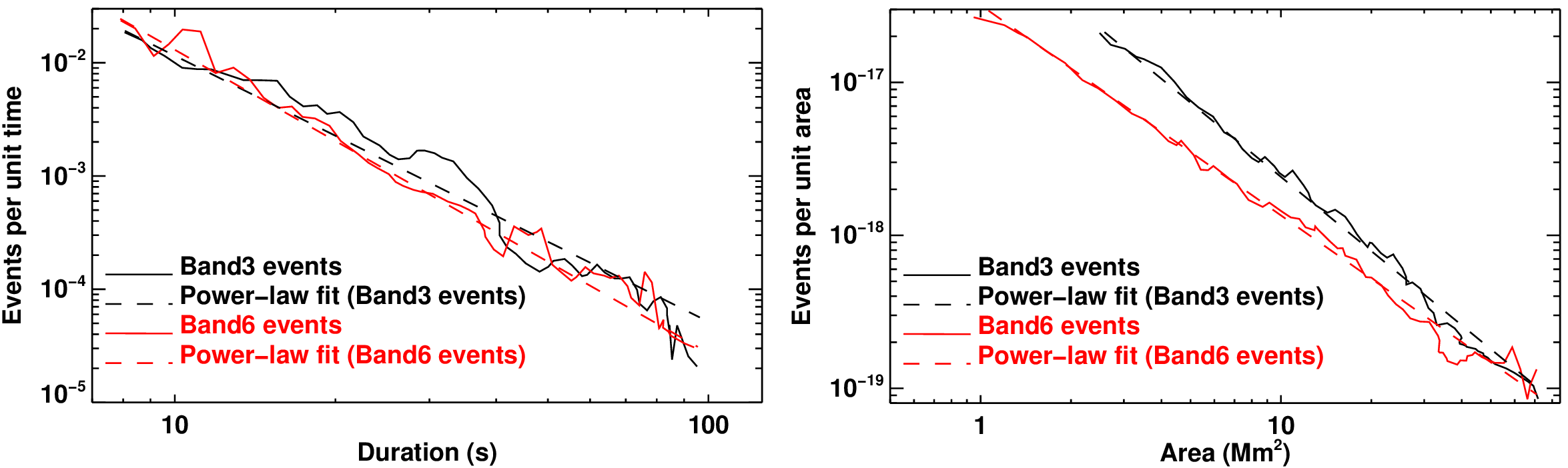}
\caption{Left panel: frequency distribution of the duration of the transient brightenings detected at 1.26\,mm and 3\,mm (red and black solid curves, respectively). The dashed red and dashed black lines show power-law fits of those distributions. Right panel: same as left panel for the frequency distribution of maximum area.}
\label{fig:histo}
\end{figure*}

Using the scheme presented by Nindos et al. (2018) we divided the
pixels of  each image into two groups, network pixels and
intra-network pixels (see  Alissandrakis et al. 2020 for a detailed
discussion about possible pixel segregation schemes). At 3\,mm the
situation is similar to that reported in Paper II, that is, there is a
rather weak tendency (73\%, see Table \ref{tab:prop})  for the
transient brightening pixels to appear at the boundaries of the
network rather than in intra-network regions. The situation is somehow
different at 1.26\,mm where 
52\% of event pixels are not
located at network boundaries. This shows well in Fig. \ref{fig:col1}
(panel a) where we have  used red color to mark the pixels of the
transient brightenings that  occurred at 1.26\,mm during scan 1. The
situation in the other 1.26\,mm scans is similar.  Fig. \ref{fig:col1}
indicates that this trend is also shared by the events  detected at
1600 \AA\ but not by those detected at 304 \AA. The different trends
between the 1.26\,mm event locations and 3\,mm event locations can be
interpreted in terms of the different characteristics of the regions
imaged at 1 and 3\,mm; in the former much of the field of view is
occupied by the interior of a  supergranular cell which is not the
case for the latter. The diversity of observed moprhologies at 1.26 mm with
ALMA full-disk images has been demonstrated by Alissandrakis et al. (2017)
and Braj\v{s}a et al. (2018).

We performed a statistical analysis of the event maximum intensities, durations,
and areas (see Table \ref{tab:prop} and Fig. \ref{fig:histo}). At both 1.26\,mm
and 3\,mm, we do not report on the properties of the paired events because they
were practically indistinguishable from the properties of all events. The
errors attached to the mean values were derived from the standard deviations
of the corresponding parameters. The frequency distributions of the parameters
we studied were all fitted with power-law functions (see Fig. \ref{fig:histo}).
The uncertainties on their indices were derived from the propagation of
errors associated with the histogram bins assuming Poisson statistics. 

The average maximum brightness temperature of the 1.26\,mm events ranged from 
44 K to 449 K while that of the 3\,mm events ranged from 65 K 
to 511 K. The above values were measured with respect to backgrounds of
$\sim$5540-6040\,K and $\sim$7180-7500 K at 1.26\,mm and 3\,mm, respectively, after we
took into account the corrections proposed by Alissandrakis et al. (2020).
Therefore at 1.26 mm we were able to detect weaker events than at 3\,mm; the range 
of excess intensity of the latter was similar to that of the events analyzed 
in Paper II. 

\begin{figure*}[t]
\centering
\includegraphics[width=1.00\textwidth]{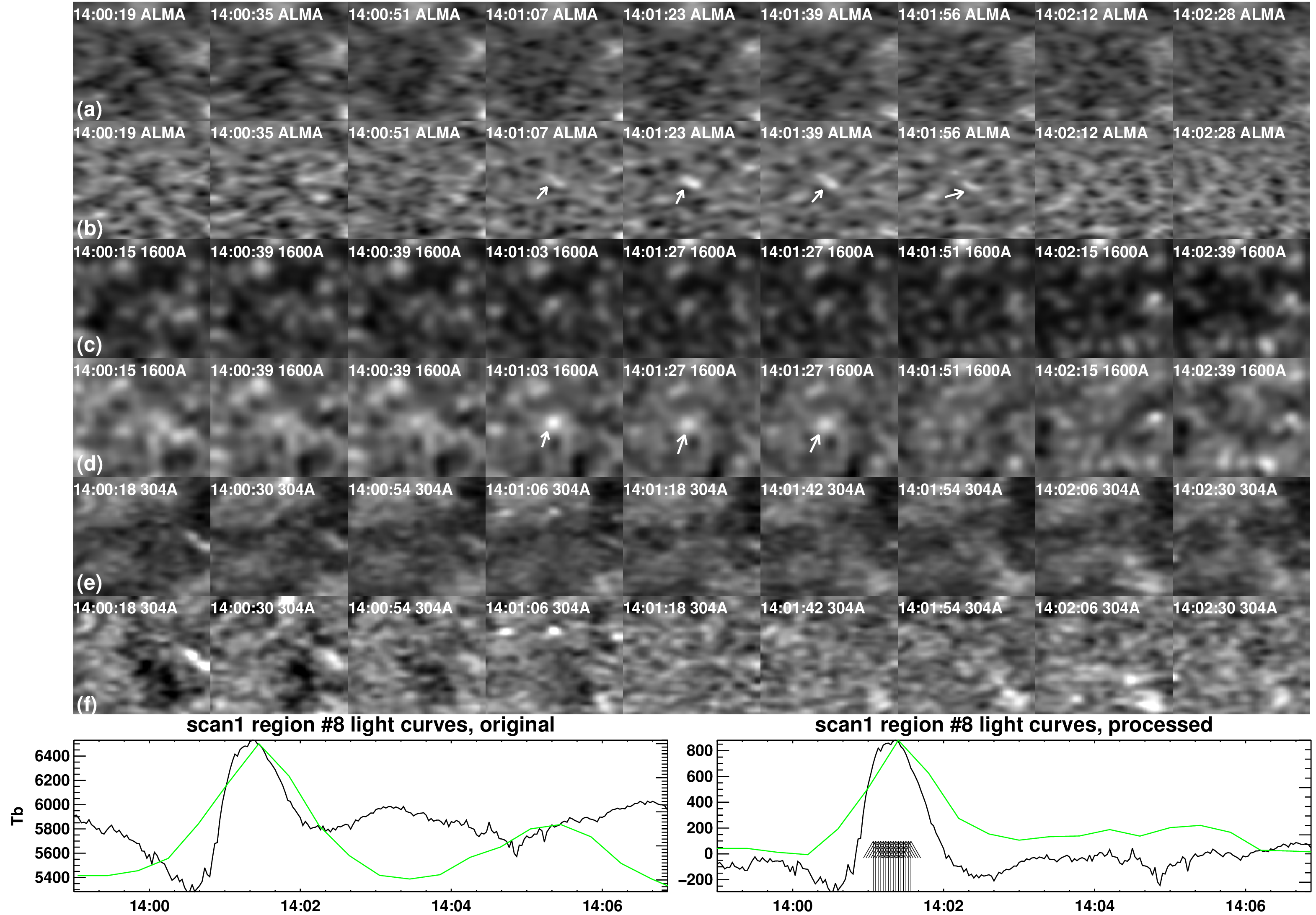}
\caption{A transient brightening 
detected both at
1.26\,mm and 1600 \AA\ scan 1 data. In rows (a) and (b) we show representative 1.26\,mm images before and after the subtraction of the average 1.26\,mm image. Rows (c)-(d) and (e)-(f) have the same format as rows (a)-(b) but they show the corresponding 1600 \AA\ data and 304 \AA\ data, respectively. The white arrows point to the transient brightening in the 1.26\,mm and 1600 \AA\ data. The field of view of all images is $20\arcsec
\times 20\arcsec$. In the bottom row we present the event time profiles at 1.26\,mm (black curves) and 1600 \AA\ (green curves) before and after our processing (left and right panels, respectively). The black arrows show the times in which the 1.26\,mm emission is greater than the 2.5$\sigma$ threshold above average. The 1600 \AA\ time profiles have been normalized to match the
vertical range of the 1.26\,mm plots.}
\label{fig:case1}
\end{figure*}

The spatial resolution of the ALMA data sets affected the mean maximum
areas of the events that we measured. The superior 1.26\,mm spatial
resolution allowed the  detection of smaller events at 1.26\,mm than at
3mm. Furthermore the average maximum size of the 3\,mm events was
smaller than the one reported in  Paper II due to the better spatial
resolution of the 2018 data. The duration  of events (quantified by
their FWHM) in all  data sets was
similar. We also note that all events were of the gradual rise and
fall type (see Fig. \ref{fig:case1} for an example) suggesting a
thermal origin.

In Fig. \ref{fig:case1} we show one characteristic event that was detected at
both 1.26\,mm and 1600 \AA. Due to their weak intensity all events at
both 1.26\,mm 
and 3\,mm were not readily visible in the plain images; their visual 
identification was made possible after we subtracted the average image from 
each datacube image. In these difference images the 3\,mm events appear as 
unresolved bright patches while some of the 1.26\,mm events are marginally resolved
(for example, see the event of Fig. \ref{fig:case1}). In the same figure the
emission from the transient brightening stands out clearly from the residuals
of the oscillations.

\subsection{Energy budgets of transient brightenings}

For the calculation of the energy budgets of the transient brightenings we
assume that they emit thermal free-free radiation, the electron density and 
apparent volume do not change significantly in the course of the events, and
that the filling factor is unity. Then we follow the method outlined in Paper 
II. Briefly, the energy is proportional to: (1) The maximum excess electron 
temperature of a given event above background which is equal to the maximum 
excess brightness temperature above background  because the events are 
optically thick. (2) The electron density which is obtained from the values
of the electron densities tabulated in the Fontenla et al. (1993) FAL C model
for the heights that correspond to the range of electron temperatures associated with the 1.26\,mm and 3\,mm emission (see also Alissandrakis et al. 2017; 2020 for
details). (3) The apparent volume which is derived from the maximum area of the
events (see Section 4.2) assuming their height is equal to their horizontal
size. 

The results of our computations appear in Table \ref{tab:energ}. For comparison
in the same table we also give the results from Paper II. At 1.26\,mm the energy
of the transient brightenings ranges from $1.8 \times 10^{23}$ erg to $1.1 
\times 10^{26}$ erg while at 3\,mm it ranges from $7.2 \times 10^{23}$ erg to 
$1.7 \times 10^{26}$ erg. Therefore at 3\,mm we derived smaller values for the
lower energies of the events than in  Paper II (difference of almost a factor 
of two), probably due to the detection of smaller and somehow weaker events 
(see section 4.2). The same arguments apply for the interpretation of the even 
lower energies detected at 1.26\,mm (a factor of four smaller than at 3\,mm). 

\begin{table*}
\begin{center}
\caption{Energetics of the 1.26\,mm and 3\,mm transient brightenings}
\begin{tabular}{lcccc}
\hline
Data set          & Minimum energy  & Maximum energy  & Power-law index & Power per unit area \\
                 & ($10^{23}$ erg) & ($10^{26}$ erg) &                 & ($10^4$ erg cm$^{-2}$ s$^{-1}$) \\
\hline
April 2018, 1.26\,mm & 1.8 $\pm$ 0.1   & 1.1 $\pm$ 0.3 & 1.64 $\pm$ 0.04 & 1.1 \\
April 2018, 3\,mm & 7.2 $\pm$ 0.2   & 1.7 $\pm$ 0.3 & 1.73 $\pm$ 0.06 & 1.8 \\
March 2017, 3\,mm & 15.0 $\pm$ 1.0  & 1.0 $\pm$ 0.2 & 1.67 $\pm$ 0.05 & 1.9 \\
(Paper II)       &                 &               &                   \\      
\hline
\label{tab:energ}
\end{tabular}
\end{center}
\end{table*}

The range of calculated energies fall within the range of values reported in
previous publications that discuss quiet Sun transient brightenings. This was 
already mentioned in Paper II for the energies associated with the March 2017 
3\,mm events. The new finding here is that the low end values of the energies 
of the 1.26\,mm events is almost one order of magnitude smaller than the presumed high-end 
energy limit of nanoflares (10$^{24}$ erg) and among the lowest energies 
ever reported. Publications that have reported the detection of events with
energies below 10$^{24}$ erg include Aschwanden et al. (2000; $5 \times 10^{23}$
erg), Parnell \& Jupp (2000; 10$^{23}$ erg) and Subramanian et al. (2018; $3
\times 10^{23}$ erg). 

\begin{figure}[t]
\centering
\includegraphics[width=0.50\textwidth]{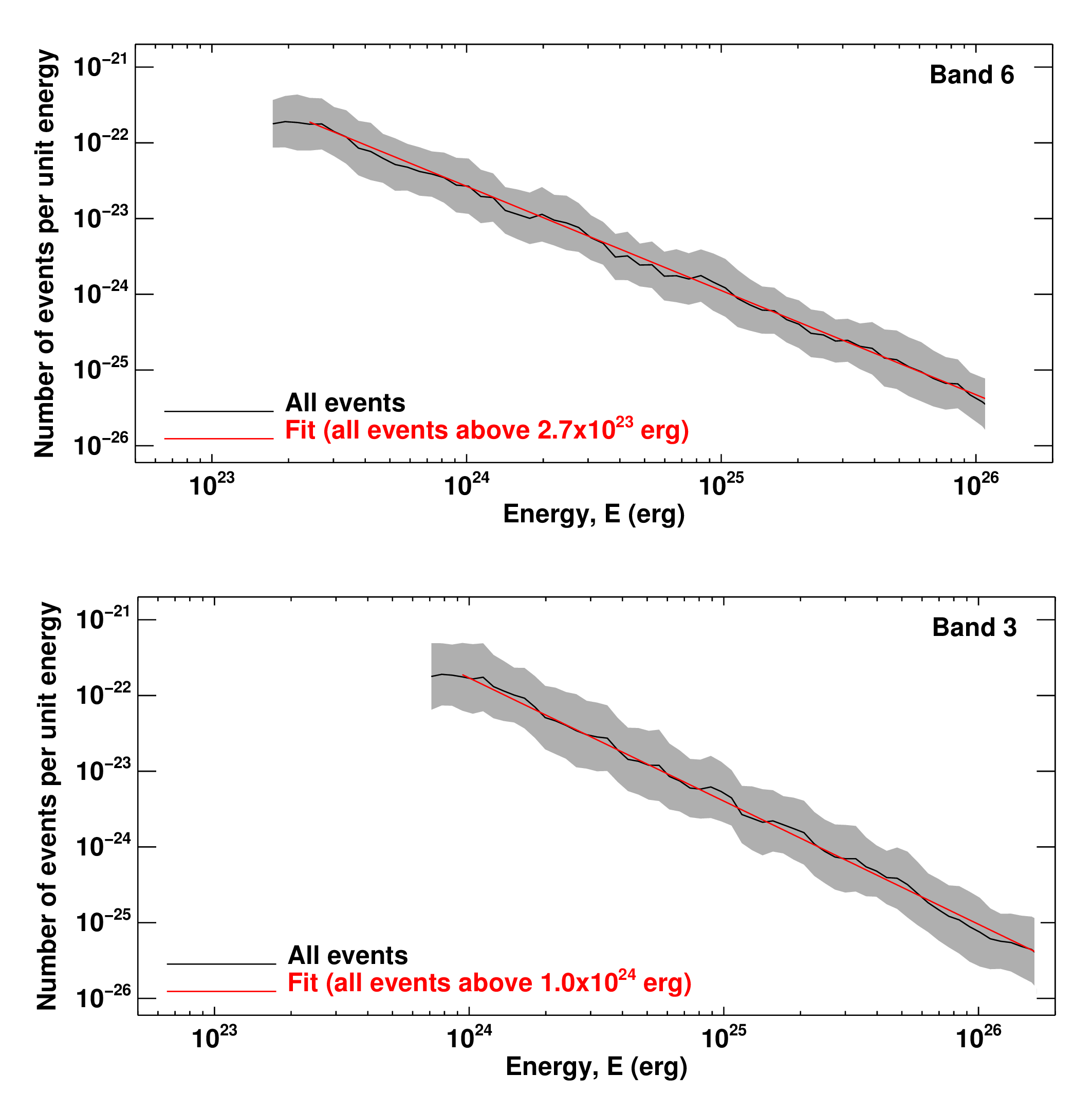}
\caption{Top panel: The black curve shows the frequency distribution of the energy of the transient brightenings detected at 1.26\,mm. The gray band denotes the uncertainties (see text for details). The red line shows the power-law  fit, with index of 1.64, of the frequency distribution for energies higher than $2.7 \times 10^{23}$ erg. Bottom panel: same as top panel for the events detected at 3\,mm. The red line shows the power-law  fit, with index of 1.73, of the frequency distribution for energies higher than $10^{24}$ erg.}
\label{fig:energies}
\end{figure}

In Fig. \ref{fig:energies} we present the frequency distribution of both the 
1.26\,mm and 3\,mm transient brightenings versus their energy. The gray areas in the
figure denote the uncertainties in each frequency distribution which arise from
the energy computations and the way we assembled each frequency distribution.
If we do not take into account the low-end parts of the energies (that is, 
values lower than $2.7 \times 10^{23}$ erg and $10^{24}$ erg for 1.26\,mm and 3\,mm, 
respectively) we can fit the frequency distributions with power-law functions
that have indices of $1.64 \pm 0.04$ and $1.73 \pm 0.06$ for the 1.26\,mm and 3\,mm
events, respectively. Given the uncertainties involved, the derived power-law 
indices do not show significant differences and they are similar to that
derived for our March 2017 3\,mm data and also similar to those reported 
in previous publications about EUV and X-ray transient brightenings (see 
Paper II and references therein).

In Table \ref{tab:energ} we also give the energies per unit area and time 
(that is, power per unit area) of the transient brightenings which were 
computed after we divided the total energy of events at a given wavelength 
with the area of the field of view and the time intervals used for the 
detection of the events. A similar calculation in Paper II showed that the 
March 2017 3\,mm transient brightenings could account for about 1\% and 10\% of 
the radiative losses in the chromosphere and corona, respectively. The 3\,mm 
events analyzed in this article provide an almost identical result. 
Furthermore, it is interesting that although the occurrence rate per unit 
area of the 1.26\,mm events is higher than that of the 3\,mm events (see Section 
4.2), their power per unit area is smaller. This result could arise
from the existence of several weaker events at 1.26\,mm. 
The combined power per unit area of transient brightenings in the two 
ALMA bands was found to be $0.86 \times 10^4$ erg cm$^{-2}$ s$^{-1}$, that is,
more than a factor of two smaller than that of the 3 mm events. Therefore
the inclusion of weaker events enhances the divergence between the computed
power per unit area and the ones required for the heating of the chromosphere 
and the corona.

We attempted to put the calculations of the energetics of the oscillations 
(see Section \ref{sec:osc}) on the same footing with the calculations of the
energetics of the transient brightenings. To this end, we assumed that the 
1.26 mm field of view was covered by beam-sized oscillating elements whose
amplitude of oscillations was assumed to correspond to a brightness temperature
enhancement due to a transient brightening. Then the division of the resulting
energy of all oscillating elements with the period of oscillations and the area
of the field of view yields a crude estimate of the power per unit area 
associated with the oscillations. We found that it was $18.2 \times 10^4$
erg cm$^{-2}$ s$^{-1}$, that is a factor of 16.5 higher than the power per
unit area associated with the 1.26 mm transient brightenings. 

\section{Conclusions and summary}

Using ALMA observations of a quiet Sun region, we 
expanded our studies of chromospheric oscillations and transient 
brightenings at mm-wavelengths that were presented in Papers I 
and II, respectively, by including Band 6 
(1.26\,mm) data in our analysis. 
Compared to our previous publications, the data sets
analyzed in this article  had
better spatial and temporal resolution 
($1-2\arcsec$ and 1\,s, respectively) as well as improved frequency 
resolution in spectral power (0.5\,mHz), compared to $3-4\arcsec$, 2 s, and 1.7 mHz, 
respectively, of our 2017 observations. As in Papers I and II, 
we took particular care in separating oscillations and transients 
from each other and from background time variations and noise.

As a result of the improved frequency resolution, the spectral peaks,
in the range of about 3.6 to about 4.4 mHz, are well 
visible above background. Differences 
between the p-mode peak frequencies for Band 6 and Band 3 are
within the spectral resolution of the analyzed power spectra.
The FWHM of the p-mode peaks were found in the range [3.2,3.9] mHz.
We found that short gaps in the time-series, 
inherent to ALMA observations, produce weak sidelobes to the spectral peaks 
and reduce spectral power. By ignoring the sidelobes and correcting the 
values of the spectral peak intensities, we computed the rms values of the 
oscillating component of the brightness temperature for the whole $120 
\times 120$-pixel regions employed for the study of oscillations in 
both 1.26\,mm and 3\,mm.

Given the differences between the 1.26\,mm and 3\,mm data in terms of both 
spatial resolution and field of view we also calculated: (1) the rms 
for the 1.26\,mm data after we degraded their spatial 
resolution to that of 3\,mm data, and (2) the
rms for a trimmed 3\,mm field of view equal to the 1.26 mm field of
view. The rms values from these further calculations correspond to 
data cubes of the same field of view and spatial resolution and 
hence they were appropriate for meaningful comparisons.

The p-mode oscillations in 1.26\,mm and 3\,mm exhibit small and
comparable  relative $T_{b}$ fluctuations of $\approx$ 1.7-1.8$\%$
with respect to the  average quiet Sun $T_{b}$ corresponding to rms of
137 and 109\,K, respectively. These results imply that the p-mode relative
amplitude does not appreciably change between the formation heights  of the
1.26\,mm and 3\,mm emissions.  
The energy density of the p-modes in
1.26\,mm is $\approx 3 \times {10}^{-2}
\mathrm{erg\,{cm}^{-3}}$. The $T_{b}$ fluctuations of the p-mode
oscillations  at 1.26\,mm and 3\,mm represent a significant
fraction (0.55-0.68) of the total  fluctuations (that is,
no matter being oscillatory or not, in  nature) of the
light curves; since the total fluctuations include instrumental noise, the actual fraction is still higher. 

The spatial resolution is affecting the rms of the p-mode oscillations
in both wavelengths (we also compared the 3\,mm observations of  this
work with those of Paper I). This implies that ALMA observations  of
chromospheric oscillations with a spatial resolution worse than at
least 1$\arcsec$  are not fully resolved.  One should also always bear
in mind that pixels are correlated on beam  scale which is around 18
pixels in both bands. This  suggests that at such  scales the observed
oscillations are not fully independent, and therefore some  ``mixing''
between cell and network seems inevitable.  In addition, the reduction
of the full 3\,mm field of view to its half, has a relatively small
effect in the resulting rms (factor of $\approx$ 1.1). Taking all
these effects into account we can consider the p-mode rms of 175\,K or
2.75\% of the quiet Sun, measured at 1.26\,mm (the 1.26 mm
background brightness temperature at disk center is 6343 K; see
Alissandrakis et al. 2020), as a lower limit of the actual value.

Our search for transient brightenings revealed 77 events at 1.26\,mm and 115 events
at 3\,mm. The occurrence rate per unit area of the events that we detected 
is much smaller than that of the events identified by Eklund et al. (2020); 
these authors detected 552 events in 3\,mm ALMA observations of the quiet Sun, 
which is equivalent of event occurrence rate per unit area almost two orders 
of magnitude higher than those found in both the present study and in Paper II.
One possible explanation of this discrepancy is that the quiet Sun region that Eklund et al. (2020) observed was very different from the one we did. On the other hand, it is not clear in their article that they removed the oscillations, thus some oscillation peaks may have been mistaken as transients. 

The occurrence rate per unit area of our 1.26\,mm events was higher than that of 
the 3\,mm events. This conclusion does not change even if we take into account 
differences in noise levels and spatial resolution between the two data sets 
(although there is a decrease of the relevant factor from 2.16 to 1.4). Much
of the field of view of the 1.26\,mm observations is covered by the interior of
a supergranular cell and the absence of strong background sources may have
contributed in the detection of weak events that could otherwise have been missed.

The predominance of cell pixels in the 1.26\,mm field of view may also explain 
the slight preference of the 1.26\,mm events to occur in cell regions in contrast 
to the opposite trend that was exhibited by the 3\,mm events. We note that
there is a long history of detection of transient brightenings in intranetwork
regions that dates back to the publication of such an event by Nindos et al.
(1999) to the more recent reports about the multitude of events associated 
with the cancellation of intranetwork magnetic fields (Go\v{s}i\'{c} et al.
2018). Another distinct property of the 1.26\,mm events was that they had no 
304 \AA\ counterparts although 15\% of them were paired with 1600 \AA\ events. 
Presumably the 1.26\,mm events are located at much deeper and cooler layers of 
the solar atmosphere than the 304 \AA\ events and their strength may not be
sufficient to give rise to 304\AA\ emission. 

At both wavelengths all transient brightenings show time profiles with a 
gradual rise and fall and durations of about 50 s. Furthermore, at both 
wavelengths we were able to detect events with sizes down to the size of the 
instrumental spatial resolution. At 1.26\,mm we detected
events that 
were somehow weaker than the 3\,mm  events (excess brightness temperatures of 44-449 
and 65-511 K above background, respectively). The intensity and duration of
the transient brightenings is not consistent with the relevant observed 
properties of the events discussed by Eklund et al. (2020) who attributed 
them to shock waves. 

The energies supplied by the transient brightenings were in
the ranges of $1.8 \times 10^{23}$ to $1.1 \times 10^{26}$ erg and $7.2 \times 10^{23}$ to
$1.7 \times 10^{26}$ erg for the 1.26\,mm and 3\,mm events, respectively. The 
power-law indices of the  frequency distributions of the events were similar 
(1.64 and 1.73 for the 1.26\,mm and 3\,mm, respectively) and close to the one 
reported in Paper II. At 3\,mm the lower end of the energies falls below that 
of the events of Paper II possibly due to the detection of events with smaller 
sizes. Even lower was the lower end of the energy distribution of the 1.26\,mm 
events, which is among the smallest ever reported in the literature of 
transient brightenings irrespective of the wavelength of observations.   
Although the occurrence rate per unit area of the 1.26\,mm events is higher, their 
power per unit area is smaller than that of the 3\,mm events probably due to 
the detection of many weak 1.26\,mm events. 

We also found that broadband \ha\ negative images show 
a strong similarity to the ALMA images; this is 
probably because broadband \ha\ images are dominated by the 
absorption of dark mottles/spicules/fibrils, which are located above 
network elements observed by ALMA, see Rutten
(2017) and Mart{\'\i}nez-Sykora et al. (2020). We also note
that Molnar et al. (2019) found a good correlation
between maps of  \ha\ core width and ALMA brightness temperature at  
3.0 mm.

It is well-known that the quiet choromspheric emission shows
significant time variation. Our study demonstrates that most fluctuations
at 1.26  and 3 mm are related to p-mode oscillations since they
are ubiquitous and provide more than $\sim$60\% of the total
fluctuations. The removal of oscillations from the data revealed
several transient brightenings. The combined power per unit area of
the transients in the two ALMA bands was about 230 times smaller
than the chromospheric radiative losses. Although with the inclusion
of the 1.26 mm events the  minimum observed energy was pushed to
values as low as about $2 \times 10^{23}$ erg, the amount of missing
power per unit area increased due to the occurrence of several
weaker events at 1.26 mm.

More comprehensive results both on oscillations and on the height
distribution of the occurrence rate per unit area of the transient
brightenings could be reached if: (1) snapshot images at different
spectral windows within the same ALMA band are analyzed (in our study,
data from all spectral windows of a given band were summed over), and
(2) observations at more ALMA bands become available; since late 2019,
Band 7 (0.85 mm) solar  ALMA observations have become available, and
future plans include the usage of Band 5 (1.4 mm) by the end of
2021. Furthermore studies of both oscillations and transient
brightenings will benefit significantly from the availability of data
obtained with higher spatial resolution resulting from a possible
future expansion of ALMA's long baselines.

\begin{acknowledgements}
The authors wish to thank Rob Rutten for interesting discussions on the association of ALMA 
emission and \ha. This work makes use of the following ALMA data:
ADS/JAO.ALMA2017.1.00653.S. ALMA is a partnership of ESO (representing its member states),
NSF (USA) and NINS (Japan), together with NRC (Canada) and NSC and ASIAA (Taiwan), and
KASI (Republic of Korea), in cooperation with the Republic of Chile. The Joint ALMA
Observatory is operated by ESO, AUI/NRAO and NAOJ. The use of SDO and GONG data from the
respective data bases is gratefully acknowledged.
\end{acknowledgements}

\begin{appendix}

\section{The effect of data gaps}

The observed time series, $f(t)$, can be described as the product of the original (infinite) time series, $f_0(t)$ and a window function, $W(t)$,
\be
f(t) =f_0(t) W (t) \label{f(t)}
\ee

In our case the window function is the sum of $\Pi$ functions, each of width $\Delta t_i$ (the duration of each scan) and shifted by $\delta t_i$ with respect to the start of the first scan:
\be
W(t)= \sum_{j=1}^n \Pi\left(\frac{t-\delta t_j}{\Delta t_j}\right) \label{W(t)}
\ee
where $n$ is the number of scans in a block (four for Band 6 and three for 
Band 3) and $\Pi$ is the well-known widow function:
\be
\Pi(x)=\left\{
\begin{array}{ll}
1\mbox{,~~}&|x| < 0.5 \\
           &            \\
0\mbox{,~~}&|x| > 0.5 \\
\end{array}
\right.
\ee

The observed power spectrum, $P(\nu)$, is the square of the modulus of the Fourier Transform (FT) of the observed time series:
\be
P(\nu)=\left|\tilde{f}(\nu)\right|^2 \label{psp}
\ee
where $\nu$ is the frequency and ~$\tilde{}$~ symbolizes the FT:
\be
\tilde{f}(\nu)=\int_{-\infty}^\infty f(t) e^{2\pi \nu t} dt
\ee

Using the similarity, shift and convolution theorems of FT (see, {\it e.g.\/}, Bracewell, 1965), we obtain from (\ref{f(t)}) and (\ref{W(t)}):
\be
\tilde{f}(\nu)=  \tilde{f}_0(\nu) \ast \sum_{j=1}^n  e^{-2\pi i \delta t_j \nu} \Delta t_j\, \sinc (\Delta t_j \nu)  \label{FTf3} 
\ee
where * denotes convolution and the $\sinc$ function (FT of $\Pi(t)$) is defined as:
\be
\sinc(\nu)\equiv\frac{\sin\pi \nu}{\pi \nu}
\ee

The power spectrum cannot be computed analytically from (\ref{FTf3}), unless $f_0(t)$ is specified, because it involves the module of a convolution. For the sake of illustration, let us compute the spectrum for a monochromatic oscillation of frequency $\nu_0$, whose FT is:
\be
\tilde{f}_0(\nu)=\delta(\nu-\nu_0)
\ee
$\delta$ being the Dirac delta function. In this case, substituting (\ref{FTf3}) to  (\ref{psp}), we obtain:
\be
P(\nu)=\left|\sum_{j=1}^n  e^{-2\pi i \delta t_j (\nu-\nu_0)} \Delta t_j\, \sinc [\Delta t_j (\nu-\nu_0)]\right|^2 \label{psp3}
\ee

\begin{figure}[h]
\centering
\includegraphics[width=0.95\hsize]{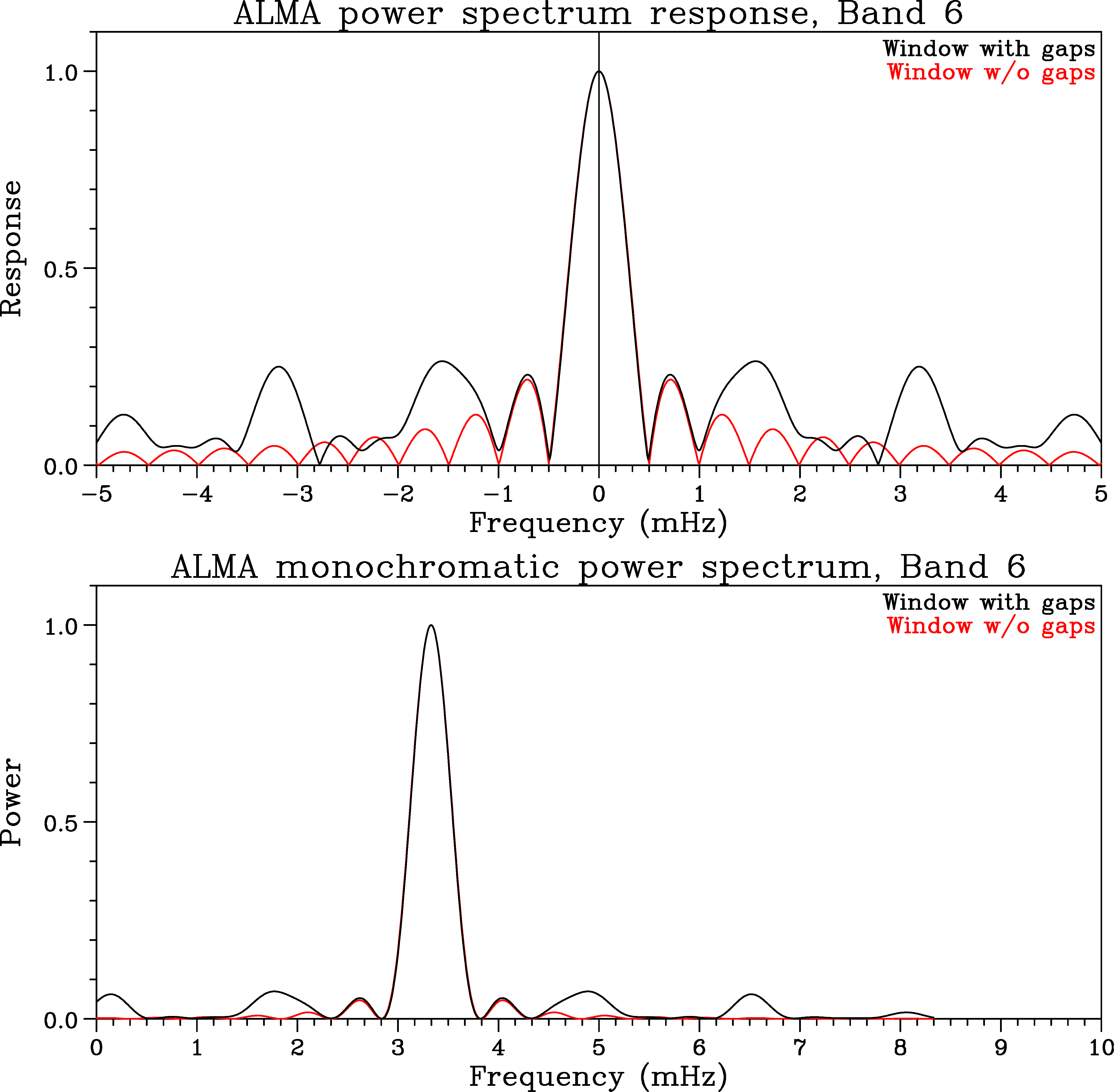}
\caption{Effect of gaps: The top panel shows the modulus of the summation term in (\ref{FTf3}) for our Band 6 observations. The power spectrum of a monochromatic oscillation at 3.333 mHz is plotted in the bottom panel. The red curve is for a 30\,min observation without gaps, the black curve is for an observation with the gaps of our Band 6 data set. All plots are normalized to a peak value of unity.}
\label{fig:gaps}
\end{figure}

An indication of the effect of gaps can be provided by the summation term in the right-hand side of (\ref{FTf3}), the modulus of which is plotted in the top panel of Fig.~\ref{fig:gaps} (black line), together with that for a data set without gaps (red line). Due to the limited total duration of the observation, both data sets show sidelobes; the gaps produce extra sidelobes, the strongest of which reach about 25\% of the peak intensity. The sidelobe effect is smaller (below 10\%) in the monochromatic power spectra, plotted in the bottom panel of the figure. 

We note that in both cases the width of the main peak is practically unaffected by the gaps. However, the value of the peak power in the presence of gaps is reduced, since, from (\ref{psp3}):
\be
P(\nu_0)=\left(\sum_{j=1}^n  \Delta t_j\, \right)^2=(\Delta t_{f}-\Delta t_{g})^2 \label{psp3a}
\ee
where $\Delta t_{f}$ is the total duration of the time series and $\Delta t_{g}$ is the duration of the gaps; thus the peak power in the series with gaps is reduced by a factor of
\be
F=\left(\frac{\Delta t_{f}-\Delta t_{g}}{\Delta t_{f}}\right)^2=\left(1-\frac{\Delta t_{g}}{\Delta t_{f}}\right)^2
\ee
with respect to the peak of the time series with gaps. In our case this factor amounts to 0.6 for Band 6 and 0.7 for Band 3.

\end{appendix}


\begin{thebibliography}{}
 
\bibitem[Alissandrakis et al.(2017)]{2017A&A...605A..78A} Alissandrakis, C.~E., Patsourakos, S., Nindos, A., \& Bastian, T.~S.\ 2017, \aap, 605, A78
\bibitem[Alissandrakis et al.(2020)]{2020A&A...640A..57A} Alissandrakis, C.~E., Nindos, A., Bastian, T.~S., et al.\ 2020, \aap, 640, A57 

\bibitem[Aschwanden et al. (2000)]{Aschwanden00} Aschwanden, M.J., Tarbell, 
T.D., Nightingale, R.W., et al.\ 2000, ApJ, 535, 1047

\bibitem[Bastian et al.(2017)]{2017ApJ...845L..19B} Bastian, T.~S., Chintzoglou, G., De Pontieu, B., et al.\ 2017, \apjl, 845, L19

\bibitem[Bracewell(1965)]{1965ftia.book.....B} Bracewell, R.\ 1965, McGraw-Hill Electrical and Electronic Engineering Series, New York: McGraw-Hill, 1965

\bibitem[Braj{\v{s}}a et al. (2018)]{2018A&A...613A..17B} Braj{\v{s}}a R., Sudar D., Benz A.~O., , et al., 2018, A\&A, 613, A17


 
\bibitem[Canfield \& Musman(1973)]{1973ApJ...184L.131C} Canfield, R.~C. \& Musman, S.\ 1973, \apjl, 184, L131 

\bibitem[Carlsson et al. (2019)]{Carlsson19} Carlsson, M., De Pontieu, B., \& Hansteen, V.H. 2019, ARA\&A, 57, 189

\bibitem[da Silva Santos et al.(2020)]{2020A&A...643A..41D} da Silva Santos, J.~M., de la Cruz Rodr{\'\i}guez, J., White, S.~M., et al.\ 2020, \aap, 643, A41

\bibitem[de Wijn et al. (2009)]{dewijn09} de Wijn, A. G., McIntosh, S. W., \& De Pontieu, B. 2009, ApJ, 702, L168

\bibitem[Eklund \emph{et al.}(2021)]{2021RSPTA.37900185E}Eklund, H., Wedemeyer, S., Snow, et al. 2021, Philosophical Transactions of the Royal Society of London Series A, 379, 20200185

\bibitem[Eklund et al.(2020)]{2020A&A...644A.152E} Eklund, H., Wedemeyer, S., Szydlarski, M., et al.\ 2020, \aap, 644, A152

\bibitem[Fontenla et al. (1993)]{fontenla93} Fontenla, J. M., Avrett, E. H., \& Loeser, R. 1993, ApJ, 406, 319

\bibitem[Go\v{s}i\'{c} et al. (2018)]{gosic18} Go\v{s}i\'{c}, M., de la Cruz Rodr{\'\i}guez, J., De Pontieu, B., et al. 2018, ApJ, 857, 48

\bibitem[Guevara G{\'o}mez et al.(2021)]{2021RSPTA.37900184G} Guevara G{\'o}mez, J.~C., Jafarzadeh, S., Wedemeyer, S., et al.\ 2021, Philosophical Transactions of the Royal Society of London Series A, 379, 20200184

\bibitem[Henriques et al. (2016)]{henriques16} Henriques, V. M. J., Kuridze, D., Mathioudakis, M., \& Keenan, F. P. 2016, ApJ, 820, 124

\bibitem[Jafarzadeh et al.(2019)]{2019A&A...622A.150J} Jafarzadeh, S., Wedemeyer, S., Szydlarski, M., et al.\ 2019, \aap, 622, A150
\bibitem[Jafarzadeh et al.(2021)]{2021RSPTA.37900174J} Jafarzadeh, S., Wedemeyer, S., Fleck, B., et al.\ 2021, Philosophical Transactions of the Royal Society of London Series A, 379, 20200174
\bibitem[Jess et al.(2015)]{2015SSRv..190..103J} Jess, D.~B., Morton, R.~J., Verth, G., et al.\ 2015, \ssr, 190, 103
\bibitem[Lemen et al.(2012)]{2012SoPh..275...17L} Lemen, J.~R., Title, A.~M., Akin, D.~J., et al.\ 2012, \solphys, 275, 17

\bibitem[Loukitcheva (2019)]{loukitcheva19} Loukitcheva, M. 2019, Advances in Space Research, 63, 1396
\bibitem[Loukitcheva et al.(2006)]{2006A&A...456..713L} Loukitcheva, M., Solanki, S.~K., \& White, S.\ 2006, \aap, 456, 713
\bibitem[Loukitcheva et al.(2019)]{2019ApJ...877L..26L} Loukitcheva, M.~A., White, S.~M., \& Solanki, S.~K.\ 2019, \apjl, 877, L26
\bibitem[Mart{\'\i}nez-Sykora et al.(2020)]{2020ApJ...891L...8M} Mart{\'\i}nez-Sykora, J., De Pontieu, B., de la Cruz Rodr{\'\i}guez, J., et al.\ 2020, \apjl, 891, L8
\bibitem[Molnar et al.(2019)]{2019ApJ...881...99M} Molnar, M.~E., Reardon, K.~P., Chai, Y., et al.\ 2019, \apj, 881, 99


\bibitem[Nindos et al.(1999)]{Nindos99} Nindos, A., Kundu, M.R., White, S.M. 
1999, ApJ, 513, 983 
\bibitem[Nindos et al.(2018)]{2018A&A...619L...6N} Nindos, A., Alissandrakis, C.~E., Bastian, T.~S., et al.\ 2018, \aap, 619, L6 
\bibitem[Nindos et al.(2020)]{nindos20} Nindos, A., Alissandrakis, C.~E., Patsourakos, S., et al.\ 2020, \aap, 638, A62 (Paper II)
\bibitem[Parnell \& Jupp (2000)]{Parnell00} Parnell, C.E. \& Jupp, P.E. 2000,
ApJ, 529, 554
\bibitem[Patsourakos et al.(2020)]{2020A&A...634A..86P} Patsourakos, S., Alissandrakis, C.~E., Nindos, A., et al.\ 2020, \aap, 634, A86 (Paper I)

\bibitem[Rutten(2017)]{2017A&A...598A..89R} Rutten, R.~J.\ 2017, \aap, 598, A89

\bibitem[Schmieder(2007)]{2007AIPC..895...49S}Schmieder, B.: 2007, {\it Fifty Years of Romanian Astrophysics}, 895, 49

\bibitem[Shibasaki et al. (2011)]{Shibasaki et al. 2011}
Shibasaki, K., Alissandrakis, C.E., \& Pohjolainen, S.\, 2011, Sol. Phys., 273,
309

\bibitem[Shimizu (2015)]{Shimizu15} Shimizu, T. 2015, PhPl, 22, 101207

\bibitem[Shimojo et al.(2017a)]{2017SoPh..292...87S} Shimojo, M., Bastian, T.~S., Hales, A.~S., et al.\ 2017a, \solphys, 292, 87

\bibitem[Shimojo et al. (2017b)]{shimojo17b} Shimojo, M., Hudson, H. S., White, S. M., et al. 2017b, ApJL, 841, L5
\bibitem[Subramanian et al.(2018)]{Subramanian18} Subramanian, S., Kashyap, 
V.L., Tripathi, D., et al.\ 2018, \aap, 615, A47
\bibitem[Tsiropoula et al. (2012)]{Tsiropoula et al. 2012}
Tsiropoula, G., Tziotziou, K., Kontogiannis, I., et al.\, 2012, Space Sci. Rev.
169, 181
\bibitem[Yokoyama et al.(2018)]{2018ApJ...863...96Y} Yokoyama, T., Shimojo, M., Okamoto, T.~J., et al.\ 2018, \apj, 863, 96
\bibitem[Wedemeyer-B{\"o}hm et al.(2009)]{2009SSRv..144..317W} Wedemeyer-B{\"o}hm, S., Lagg, A., \& Nordlund, {\r{A}}.\ 2009, \ssr, 144, 317
\bibitem[Wedemeyer et al. (2016)]{Wedemeyer et al. 2016}
Wedemeyer, S., Bastian, T., Braj\v{s}a, R., et al.\, 2016, Space Sci. Rev., 200,
1

\bibitem[Wedemeyer et al. (2020)]{wedemeyer20} Wedemeyer, S., Szydlarski, M., Jafarzadeh, S., et al. 2020, A\&A, 635, A71

\bibitem[White et al.(2006)]{2006A&A...456..697W} White, S.~M., Loukitcheva, M., \& Solanki, S.~K.\ 2006, \aap, 456, 697


\end{thebibliography}
\end{document}